# Towards the Industrial Metaverse: A Game-Based VR Application for Fire Drill and Evacuation Training for Ships and Shipbuilding

Musaab H. Hamed-Ahmed
Dept. of Computer Engineering,
Universidade da Coruña
Centro de Investigación CITIC
A Coruña, Spain

Paula Fraga-Lamas
Dept. of Computer Engineering,
Universidade da Coruña
Centro de Investigación CITIC
A Coruña, Spain

Tiago M. Fernández-Caramés
Dept. of Computer Engineering,
Universidade da Coruña
Centro de Investigación CITIC
A Coruña, Spain

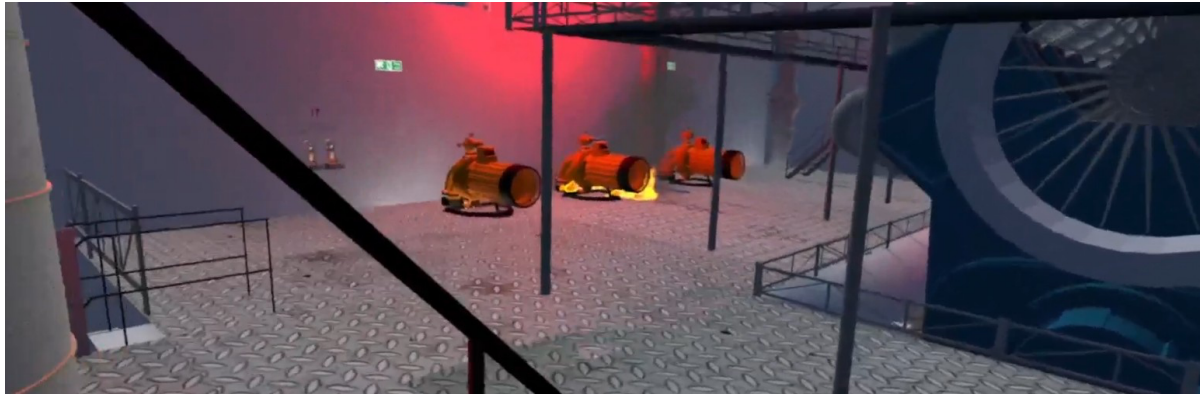

**Figure 1: Ship engine room on fire encountered by a trainee during a drill carried out with the developed application.**

## Abstract

This paper details the creation of a novel Virtual Reality-based application for the Industrial Metaverse aimed at shipboard fire emergency training for fire drill and evacuation, aligned with the Safety of Life at Sea (SOLAS) convention requirements. Specifically, the application includes gamified scenarios with different levels (e.g., varying fire intensities in engine rooms and galleys). The paper details comprehensively the VR development while providing a holistic overview and practical guidelines. Thus, it can guide practitioners, developers and future researchers to shape the next generation of Industrial Metaverse applications for the shipbuilding industry. Moreover, the paper includes the results of a preliminary user evaluation aimed at quantifying user decision-making and risk assessment skills. The presented results of the experiments provide insights into user performance and allow for pointing out at different future challenges.

This work has been funded by grants TED2021-129433A-C22 (HELENE) and PID2020-118857RA-100 (ORBALLO) funded by MCIN/AEI/10.13039/501100011033 and the European Union NextGenerationEU/PRTR, and by contract "Colaboración en proyecto ARGOS" (F22/11, CITIC).

## 1 Introduction

Fire accidents on ships pose significant risks, causing high fatalities, financial damage and environmental impact. Between 2011 and 2015, fire accidents caused 132% more fatalities than other shipboard accidents and accounted for 24% of total ship accidents [1]. To address these risks, shipbuilders integrate fire safety measures into large ship designs [2, 3] and the International Convention for the Safety of Life at Sea (SOLAS) mandates regular fire drills [4, 5].

The Industrial Metaverse [6] is a virtual environment where Industry 4.0/5.0 technologies (e.g., Extended Reality (XR), Industrial Internet of Things (IIoT), Cloud Computing and Artificial Intelligence (AI)) converge for immersive and collaborative industrial applications. It allows for real-time simulation, visualization and optimization of processes. In the maritime industry, the Industrial Metaverse is transforming practices, boosting efficiency, innovation and productivity. Despite safety advances, ship fires remain critical due to human error and insufficient training [2, 3, 7]. To prevent such issues, XR technologies offer promising solutions for immersive and interactive crew training [8, 9]. In particular, this paper presents a novel Virtual Reality (VR)-based application for the Industrial Metaverse aimed at fire drill training for ships, featuring various difficulty levels to assess trainees' decision-making and risk-assessment abilities.

## 2 Background

### 2.1 On how fire drills are carried out in ships

Fire drills are essential to ship management and overseen by the ship master [10]. As mandated by SOLAS, they must be conducted monthly, simulating realistic scenarios and focusing on vulnerable areas onboard the ship [4]. The drill sequence begins with initiating the fire alarm, followed by crew assembly at muster stations. Designated firefighters, engage in suppression and rescue operations. Reserve personnel assists as needed, and if initial efforts fail, the fixed fire extinguishing system is activated [5].

There are evaluation criteria for each step of the fire drill, assessing factors like time, performance and task completion [10]. Following drills, the ship master evaluates crew performance, providing corrective actions as needed to enhance preparedness [10].

### 2.2 XR Applications for the Maritime Industry

While XR has demonstrated effectiveness across various industries, its integration into maritime operations and shipbuilding remains limited [11, 12]. For instance, in [13] the authors identify three stages for XR application in the maritime industry which are accident prevention, response and analysis.

Augmented Reality (AR) aids maritime construction by assisting in outfitting and assembly [14, 15]. Moreover, AR applications can integrate real-time sensor measurements with platforms like Unity and Microsoft Visual Studio [16, 17]. Meanwhile, VR is used in training simulators for its cost-effectiveness and environmental benefits [8]. Existing applications include spray painting [18], safety training [19] and mooring operations [20]. Multi-role [21] and multi-player [22] simulations have also been developed for lifesaving and evacuation strategies aboard passenger ships. Moreover, prominent maritime organizations have adopted VR for training and educational initiatives [19].

### 2.3 XR Applications for Fire Drilling

The use of XR for fire drills has gained recent attention. Kang et al. [23] and Tzani et al. [24] developed AR/VR systems for fire drills, emphasizing the importance of high-quality hardware and integrating fire simulations into Unity 3D. Khan et al. [25] found that adding olfactory cues to VR fire simulations increased behavioral realism. Other authors [26, 27] explored integrating VR with AI and game engines for improved disaster evacuation training. Another project [28] developed a VR fire safety training program that significantly improved knowledge and user satisfaction. Lastly, in [29, 30] the authors highlighted the importance of standards and regulations in fire safety.

With only two notable studies, the use of XR for fire drills in shipbuilding is limited. Wu et al. [31] highlighted VR benefits in firefighting training, focusing on fire influence parameters but lacking application development and validation. Conversely, Wahidi et al. [8] integrated VR training in shipyards, showing a 14.05% increase in safety knowledge and a 68.13% cost reduction, but focused more on user alignment and gestures than on detailed fire drill applications.

### 2.4 Analysis of the State of the Art

After reviewing the state of the art, it is clear that XR applications in the maritime industry are scarce and nascent. The academic literature is limited, with few comprehensive journal articles and mostly outdated conference papers lacking detailed VR development information. Fire drilling applications are mainly designed for other scenarios, with only two targeting ship fire drills in engine rooms, none of which are game-based. The game-based proposal proposed in this paper offers hands-on development to guide researchers in creating effective training tools for ships within current regulatory frameworks. Specifically, this paper includes the following contributions:

- An overview of XR applications for the maritime industry.
- An overview of the main characteristics of the most recent XR fire drilling systems
- An analysis of the most relevant research on VR-based fire drilling for ships and shipbuilding.
- A description of the design of a game-based learning VR application, aiming to improve safety knowledge acquisition.
- Guidelines for VR development, guiding the next generation of Industrial Metaverse applications in shipbuilding.
- A preliminary user evaluation, offering insights into the application efficiency and future challenges.

## 3 Design of the System

### 3.1 Selected Regulatory Framework for Fire Drills

SOLAS, established in 1974 and amended over time, is the cornerstone of maritime safety and environmental protection regulations, covering eleven chapters on various aspects of maritime safety [32].

For fire safety, SOLAS mandates regulations on fire detection, extinguishing, Personal Protective Equipment (PPE), evacuation procedures and handling dangerous materials [4]. Key regulations include the presence of visual and audible alarms (Regulation II-2/7 and III 6.4.2), proper evacuation routes and muster areas (Regulation II-2/12), firefighting equipment, crew training and fire containment protocols (Regulation II/2 2.1.7, 5, 7.5.1, 15.2.1.1, 15.2.3, 16.2, 18.8, and III 35). They also mandate regular fire drills for crew familiarity with evacuation procedures (Regulation II-2/15.2.2 and III/19.3) and proper signage on escape routes (Regulation II/13.3.2.5).

### 3.2 Design of the Virtual Scenario

Fires occur more often in the engine room and galley due to machinery, oil and electrical faults [2, 3]. Therefore, these areas were chosen for scenario development. The objective is to create an immersive training experience for crew members to simulate fire emergencies, evaluating their response and risk analysis skills. Insights from fire drill scenarios [10, 5] and a training booklet [33] inform the design, excluding an automatic fire detection system. Participants navigate freely in the engine room or galley, identifying the fire location using auditory (burning sounds) or visual (flames) cues. Upon detection, they inform the ship master and activate the fire alarm. They then assess the fire severity to determine if it is controllable or possesses an imminent threat. Finally, they evacuate to the designated muster area.

The developed VR-based Industrial Metaverse application uses gamification principles [34, 35, 36] to enhance problem-solving skills and learning. Trainees progress through stages with varying fire intensity levels and locations, effectively evaluating their performance.

## 3.3 Selected Hardware and Software

The Meta Oculus Rift S Head-Mounted Display (HMD) was chosen for developing the application, featuring inside-out tracking and 6 Degrees of Freedom (DoF) [37]. Unity Editor version 2022.3.15f1 was used as game engine with Universal Rendering Pipeline (URP), while the XR Interaction Toolkit facilitated immersive interactions.

OpenXR managed the communications between the Oculus HMD and the application for cross-platform compatibility [38]. Blender was used alongside Unity for parts of the virtual environment due to its robust features and VR readiness [39].

## 4 Implementation

### 4.1 Scenario levels

To construct the scenes, four main scenes were identified, each corresponding to a level in the application. These scenes were determined by the location (galley or engine room) and the presence of an extinguishable fire. Initially, in Level 1 trainees handle a small galley fire to learn basic fire management techniques. Level 2 increases fire intensity with an inextinguishable blaze, testing adaptability and decision-making. Level 3 features an engine room fire, requiring reliance on safety protocols without explicit instructions, but it remains extinguishable. Level 4 presents the greatest difficulty with a rapidly intensifying, inextinguishable fire in the engine room, challenging trainees' firefighting skills and resilience.

### 4.2 Virtual Environment

The development process involved importing a cargo ship model and constructing the ship interior cabin using Blender and the HomeBuilder add-on, inspired by a tween-decker general cargo ship arrangement. Figure 2 shows the developed galley area within the virtual environment.

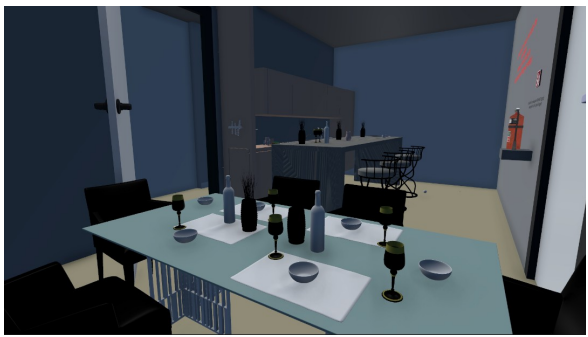

**Figure 2: Galley area of the ship.**

For the engine room, conceptual designs were utilized, and a VR-compatible engine model was imported and modified (the final model is shown in Figure 3). Surrounding elements like ladders, fences and machinery were constructed. Moreover, signage denoting fire safety measures were added.

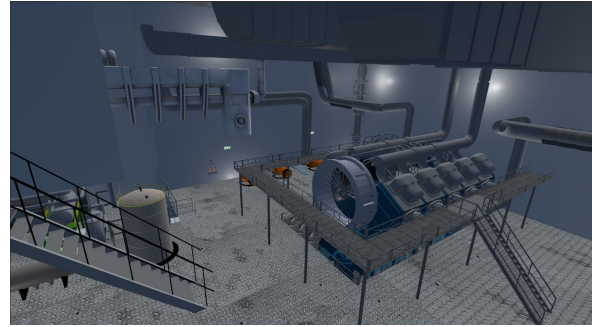

**Figure 3: Engine room of the ship.**

Interactive elements such as fire extinguishers, alarms and emergency phones were included. Simulated fire and smoke effects were integrated using Unity's particle system. Figure 4 shows the implemented intractability of the fire extinguisher with the player's hand and the simulated fire and fire extinguisher agent.

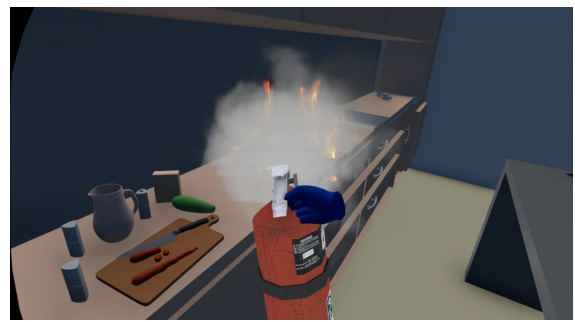

**Figure 4: Instant when the trainee is extinguishing a fire.**

### 4.3 UI implementation

The application's User Interface (UI) is designed in a dedicated scene as the initial interaction point for trainees upon launch. Using Unity XR features, the UI Canvas includes text elements for clarity and button interactions with visual feedback. The UI has four main sections. To ensure visual coherence between the trainee's position and UI menus, the Ray Light Interaction is used. This integrates a ray interactor for controllers, enabling effective interaction with options. Parameters like ray length and color are adjusted for optimal visibility as illustrated in Figure 5.

## 5 Experiments

### 5.1 Preliminary User Evaluation

User evaluation tests assessed trainees' performance based on time spent on each level. Ten students from the Erasmus Mundus Master Program on Sustainable Ship and Shipping 4.0 participated, reporting their familiarity with fire drills, VR and video games

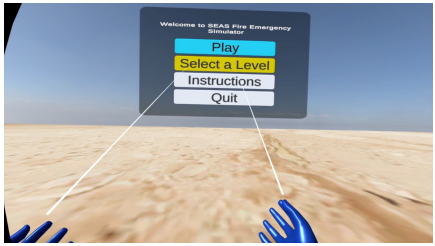

**Figure 5: Main menu.**

(summarized in Table 1). It was assumed that those with more experience in gamified interactions would be more agile in navigating and interacting with the VR application. The testers were guided on using the controllers and interacting with the UI and application objects, and all followed the fire drill instructions on the application UI.

| Tester No. | Experience with Fire Drills | Experience with VR | Experience with Playing Video Games |
|---|---|---|---|
| 1 | High | High | High |
| 2 | High | Low | Low |
| 3 | Low | Low | High |
| 4 | Medium | Low | Low |
| 5 | Low | Low | High |
| 6 | Low | Low | High |
| 7 | Low | Low | Low |
| 8 | High | Low | High |
| 9 | Low | Low | Low |
| 10 | High | Medium | High |

**Table 1: Previous experience of the testers.**

Figure 6 shows the time spent on each level by every tester. Level 1 is the most time-consuming due to the lengthy fire extinguishing process, which takes 45 seconds (28 seconds longer than Level 3). Additionally, testers needed extra time to explore the cabin area and engine room in Levels 1 and 3 due to the level design. In contrast, Levels 2 and 4 had no extinguishing process, and trainees were already familiar with the virtual scenario. As the levels progressed, trainees moved faster and understood the fire drill better, indicating that the gamified approach effectively facilitates rapid learning of the SOLAS procedure.

Another notable factor is the significant spikes in time spent, particularly in Level 1. According to the data on Table 1, the testers with low experience in VR and video games took longer to familiarize themselves with the controls and virtual environment. In contrast, those with more VR or gaming experience used the application more easily and intuitively, confirming the pre-experiment assumption.

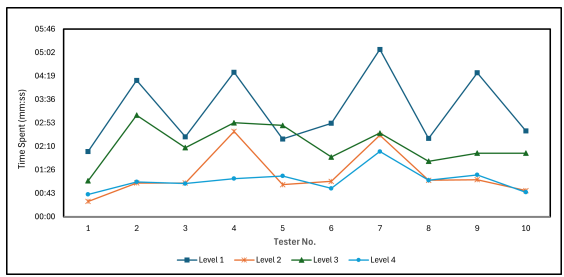

**Figure 6: Time spent on each level per tester.**

If we take as a reference user 1 (who had significant experience in fire drills, VR and video gaming) and only consider the testers' experience in video gaming (ignoring VR experience), a clear correlation emerges: experienced gamers have significantly lower time differences compared to user 1 across all levels than those with less gaming experience. Specifically, the slowest experienced gamers required 52, 39, 102 and 34 seconds more than user 1 to complete levels 1 to 4, respectively. In contrast, the slowest non-experienced gamers required 188, 129, 121 and 79 seconds more for the same levels, averaging more than double the time of experienced gamers.

Finally, for task familiarity and completion, all trainees followed instructions and made correct decisions at each level, except testers 4, 8 and 9. Specifically, testers 4 and 9 attempted to extinguish the fire in Level 2, while tester 8 evacuated before extinguishing the fire in Level 3. This indicates the need for a real-time task tracker to guide trainees if they miss a specific task. Overall, most testers enhanced their decision-making skills in assessing fire severity and determining whether to extinguish the fire or to evacuate.

The experiments also highlighted areas for future improvement of the VR application. More guidance is needed, as two testers made mistakes in decision-making and risk assessment regarding whether to extinguish the fire. Additionally, one tester skipped the extinguishing process, indicating the need for an extra assessment information layer.

In addition, it must be indicated that some of the testers expressed some frustration concerning motion sickness, and one tester even had to pause the experiment. This is essentially related to the use of the selected VR HMD, but it should be also considered by future developers and researchers.

# 6 Conclusions

This paper described the development and preliminary evaluation of a game-based VR application for the future Industrial Metaverse. Specifically, it is aimed at training ship crew members and ship-building operators on fire drills and evacuations. The application trains crew members in crucial responses to fire incidents, including fire discovery, reporting, alarm activation, and decision-making regarding fire suppression or evacuation.

Performance tests showed consistent operation with minimal frame rate fluctuations, demonstrating the application suitability for VR training. In addition, a preliminary validation with ten users indicated that those with video gaming experience completed fire drill tasks faster than those without.

Future research will address issues detected during preliminary evaluation, such as adding a map menu or a quick tour over the scenarios, intuitive overlays for better navigation, and real-time task tracking to guide the trainees if they missed a specific task. Further development will enhance realism through advanced rendering pipelines, detailed ship modeling and solutions for motion sickness. In addition, integrating multiplayer functionality, AI-controlled characters and Computational Fluid Dynamics (CFD) fire simulation models hold promise for further enriching the training experience.